\documentclass[onecolumn,prc,superscriptaddress,unsortedaddress,preprintnumbers,amsmath,amssymb]{revtex4}

\usepackage{graphicx}
\usepackage{dcolumn}
\usepackage{bm}
\usepackage{CJK}
\usepackage{CJKulem}
\usepackage{txfonts}
\usepackage{bm}
\usepackage[dvipdfmx,bookmarks=true,colorlinks,%
citecolor=blue,linkcolor=blue,anchorcolor=blue,filecolor=blue,urlcolor=blue,%
           ]{hyperref}

\def\beq{\begin{equation}}
\def\eeq{\end{equation}}
\def\bea{\begin{eqnarray}}
\def\eea{\end{eqnarray}}

\def\fun#1#2{\lower3.6pt\vbox{\baselineskip0pt\lineskip.9pt
  \ialign{$\mathsurround=0pt#1\hfil##\hfil$\crcr#2\crcr\sim\crcr}}}
  
\begin{document}
\preprint{}

\title{Neutron star cooling with a dynamic stellar structure}

\author{J. M. Dong}\email[ ]{dongjm07@impcas.ac.cn}\affiliation{Institute of Modern Physics,
Chinese Academy of Sciences, Lanzhou 730000, China}
\author{L. J. Wang}
\affiliation{Department of Physics and Astronomy, University of
North Carolina, Chapel Hill, North Carolina, 27516-3255, USA}
\author{W. Zuo}
\affiliation{Institute of Modern Physics, Chinese Academy of
Sciences, Lanzhou 730000, China} \affiliation{School of Physics,
University of Chinese Academy of Sciences, Beijing 100049, China}

\date{\today}

\begin{abstract}
The observations combined with theory of neutron star (NS) cooling
play a crucial role in achieving the intriguing information of the
stellar interior, such as the equation of state (EOS), composition
and superfluidity of dense matter. The traditional NS cooling theory
is based on the assumption that the stellar structure does not
change with time. The validity of such a static description has not
yet been confirmed. We generalize the theory to a dynamic treatment;
that is, continuous change of the NS structure (rearrangement of the
stellar density distribution with the total baryon number fixed) as
the decrease of temperature during the thermal evolution, is taken
into account. It is found that the practical thermal energy used for
the cooling is slightly lower than that is estimated in static
situation, and hence the cooling of NSs is accelerated
correspondingly but the effect is rather weak.
Therefore, the static treatment is a good approximation in the
calculations of NS cooling.
\end{abstract}

\maketitle
\noindent{} Online-only material: color figures \maketitle

\section{Introduction}\label{intro}\noindent
Neutron stars (NSs) as a type of compact objects, contain matter of
supranuclear density in their interiors which are not accessible in
modern laboratory experiments. They have a lot of extreme and
intriguing features that are unique in the universe, including rapid
rotation, extremely strong magnetic field, superstrong gravitation,
superfluidity and superconductivity inside, superprecise spin period
(Haensel et al. 2006). Accordingly, as natural laboratories, they
involve superrich physics that are related to various branches of
current physics and astronomy, and hence they greatly promote the
development of fundamental physics in extreme environments. At
present, many properties can be measured with increasingly improved
accuracies, such as the mass, radius, spin period and its derivative
with respect to time, glitch, giant flares, quasi-periodic
oscillations and surface magnetic field, which help one to grasp
important information on NSs. However, very little knowledge of NS
interior can be achieved solidly through these observations. The NS
thermal evolution provides a possibility to study its interior
physics and some difficult issues in nuclear physics such as the
equation of state (EOS) of supranuclear densities (Page et al.
2004).

The rapid cooling of the NS in Cassiopeia A was reported from an
analysis of several Chandra observations (Heinke \& Ho 2010). It has
ignited a great interest in the exploration of NS thermal evolution,
and thus numerous developments have performed on the theoretical
models that are used to interpret these observational data (Blaschke
et al. 2012, 2013; Bonanno et al. 2014; Newton et al. 2013; Noda et
al. 2013; Page et al. 2011; Shternin et al. 2011; Sedrakian 2013;
Yang et al. 2011). Owing to the well-known age (Fesen et al. 2006)
and well-measured surface temperature for ten years (Heinke \& Ho
2010), this NS serves as a valuable opportunity to explore the
knowledge of the neutron star matter at high densities. For
instance, the observed rapid cooling was interpreted as the
triggering of enhanced neutrino emission resulting from the neutron
$^3P_2$ pairing in the NS core, and it was claimed that such rapid
cooling is the first direct evidence that superfluidity and
superconductivity occur at supranuclear densities within NSs (Page
et al. 2011). Quantitatively, the superfluidity gap of neutron
$^3P_2$ channel is found to be around 0.1 MeV (Page et al. 2011).
However, another group reported that a statistically significant
temperature drop is not seen for the NS in Cassiopeia A (Posselt et
al. 2013). And also a microscopic calculation gives a small $^3P_2$
pairing, which differs from the result of D. Page et al (Dong et al.
2013; Dong et al. 2016). Anyway, reliable observations of NS thermal
evolution provide a powerful probe to grasp information on the NS
interior.

As a consequence of improved measurements of thermal emission from
cooling NSs, it has become clear that the observations cannot be
explained on the basis of a single universal cooling curve (Yakovlev
\& Pethick 2004). Thus, a reliable theory for the NS cooling is
indispensable to predict accurately the evolution of the NS surface
temperature, and to gain the important information about the stellar
interior in combination with observations. At present, the structure
of a given isolated NS is believed to not change with time in all
the previous investigations of thermal evolution, referred to as the
static treatment. In view of the great importance of the NS cooling
in both astrophysics and nuclear physics, in this work, the previous
static approach to describe the NS cooling is generalized to a
dynamic one; that is, the temperature-dependent (and hence the
time-dependent) change of the stellar structure during the cooling
is included.

\section{EOS of dense matter at finite temperature}\label{intro}\noindent

It is necessary to establish a NS structure for the calculation of
NS cooling. The EOS of NS matter as the input for the building of
stellar structure is obtained from the relativistic mean field (RMF)
theory in which the temperature effect can be readily introduced.
For the dense matter made of nucleons ($B$=p,n) and leptons ($l$=e,
$\mu$), the total interacting Lagrangian density is given by
\begin{eqnarray}
\mathcal{L} &=&\overline{\psi }_{B}(i\gamma ^{\mu }\partial _{\mu
}-M-g_{\sigma }\sigma -\frac{g_{\rho }}{2}\gamma ^{\mu }{\bm\tau }\cdot {\bm%
\rho _{\mu }+g_{\omega }\gamma ^{\mu }\omega _{\mu })}\psi _{B}  \notag \\
&&+\frac{1}{2}\partial _{\mu }\sigma \partial ^{\mu }\sigma -(\frac{1}{2}%
m_{\sigma }^{2}\sigma ^{2}+\frac{1}{3}g_{2}\sigma ^{3}+\frac{1}{4}%
g_{3}\sigma ^{4})  \notag \\
&&-\frac{1}{4}\Omega _{\mu \nu }\Omega ^{\mu \nu
}+\frac{1}{2}m_{\omega
}^{2}\omega _{\mu }\omega ^{\mu }-\frac{1}{4}{\bm R}_{\mu \nu }\cdot {\bm R}%
^{\mu \nu }+  \notag \\
&&\frac{1}{2}m_{\rho }^{2}{\bm\rho }_{\mu }\cdot {\bm\rho }^{\mu }+\frac{%
\zeta }{4!}g_{\omega }^{4}(\omega _{\mu }\omega ^{\mu })^{2}+\Lambda
_{v}g_{\rho }^{2}{\rho }_{\mu }\cdot {\rho }^{\mu }g_{\omega
}^{2}\omega
_{\mu }\omega ^{\mu } \notag \\
&&+\overline{\psi }_{l}(i\gamma ^{\mu }\partial _{\mu }-m_{l})\psi
_{l},
\end{eqnarray}
where $M$, $m_l$, $m_{\sigma }$, $m_{\omega }$ and $m_{\rho }$ are
the nucleon-, lepton-, $\sigma $-, $\omega$- and $\rho $-meson
masses, respectively. The field tensors for the vector meson are
given as $\Omega _{\mu\nu }=\partial _{\mu }\omega _{\nu }-\partial
_{\nu }\omega _{\mu }$ and by similar expression for $\bm {R}_{\mu
\nu}$ of $\rho $ meson. The nucleon field $\psi_{B} $ interacts with
the $\sigma ,\omega ,\rho $ meson fields $\sigma ,\omega _{\mu },\bm
\rho _{\mu }$ with the coupling constants $g_\sigma ,g_\omega$, and
$g_\rho $ respectively, and the lepton field $\psi_{l} $ is free
field. The self-coupling term of the $\sigma$ meson with coupling
constants $g_2$ and $g_3$ is responsible for reducing the
compression modulus of symmetric nuclear matter to an empirical
value (Boguta, \& Bodmer 1977). The self-coupling of omega-meson
described by the coupling constant $\zeta$, is introduced to soften
the equation of state at high density. The nonlinear mixed
isoscalar-isovector coupling described by $\Lambda_v$ modifies the
density-dependence of the symmetry energy (Fattoyev et al. 2010). In
the mean field approximation, the meson field operators are replaced
by their expectation values. There is no current in uniform nuclear
matter and thus the spatial vector components of $\omega _{\mu },\bm
\rho _{\mu }$ vanish, with only the timelike components $\omega
_{0},\bm \rho _0$ left. In addition, the charge conservation makes
sure that only the third-component of the isospin of $\rho$ meson,
i.e., $\rho _{30}$, is nonzero. In a word, the $\sigma$, $\omega
_{0}$, and $\rho _{30}$ are the nonvanishing expectation values of
meson fields in nuclear matter. The employed effective interaction
in the RMF approach is IU-FSU here, which gives a good description
of ground state properties as well as excitations of finite nuclei
(Fattoyev et al. 2010). Accordingly, the energy density
$\varepsilon$ and pressure $P$ for a zero-temperature NS matter are
given as
\begin{eqnarray}
\varepsilon &=&\underset{B=n,p}{\sum }\frac{1}{\pi ^{2}}%
\int_{0}^{k_{F,B}}dkk^{2}\sqrt{k^{2}+(m_{B}-g_{\sigma }\sigma
)^{2}}+  \notag
\\
&&\underset{l=e,\mu }{\sum }\frac{1}{\pi ^{2}}\int_{0}^{k_{F,l}}dkk^{2}\sqrt{%
k^{2}+m_{l}^{2}}+\frac{1}{2}m_{\sigma }^{2}\sigma ^{2}+\frac{1}{3}%
g_{2}\sigma ^{3}+\frac{1}{4}g_{3}\sigma ^{4} \notag \\
&&+\frac{1}{2}m_{\omega }^{2}\omega _{0}^{2}+\frac{1}{2}m_{\rho
}^{2}\rho _{30}^{2}+3\Lambda _{v}g_{\rho }^{2}\rho
_{30}^{2}g_{\omega }^{2}\omega _{0}^{2}+\frac{\zeta }{8}g_{\omega
}^{2}\omega _{0}^{4},
\end{eqnarray}
\begin{eqnarray}
P &=&\underset{B=n,p}{\sum }\frac{1}{3\pi ^{2}}\int_{0}^{k_{F,B}}dk\frac{%
k^{2}}{\sqrt{k^{2}+(m_{B}-g_{\sigma }\sigma )^{2}}}+  \notag \\
&&\underset{l=e,\mu }{\sum }\frac{1}{3\pi ^{2}}\int_{0}^{k_{F,l}}dk\frac{%
k^{2}}{\sqrt{k^{2}+m_{l}^{2}}}-\frac{1}{2}m_{\sigma }^{2}\sigma ^{2}- \notag\\
&&\frac{1}{3}g_{2}\sigma ^{3}-\frac{1}{4}g_{3}\sigma ^{4}+\frac{1}{2}%
m_{\omega }^{2}\omega _{0}^{2}+\frac{1}{2}m_{\rho }^{2}\rho _{30}^{2} \notag\\
&&+\Lambda _{v}g_{\rho }^{2}\rho _{30}^{2}g_{\omega }^{2}\omega _{0}^{2}+%
\frac{\zeta }{24}g_{\omega }^{2}\omega _{0}^{4},
\end{eqnarray}
where $k_{F,B}$ and $k_{F,l}$ are the Fermi momenta of nucleons and leptons, respectively.
The EOS based on the Haensel-Zdunik-Dobaczewski and
Negele-Vautherin (Haensel \& Zdunik 1990, Negele \& Vautherin 1973)
is applied for the NS crust.

\begin{figure}[htbp]
\begin{center}
\includegraphics[width=0.5\textwidth]{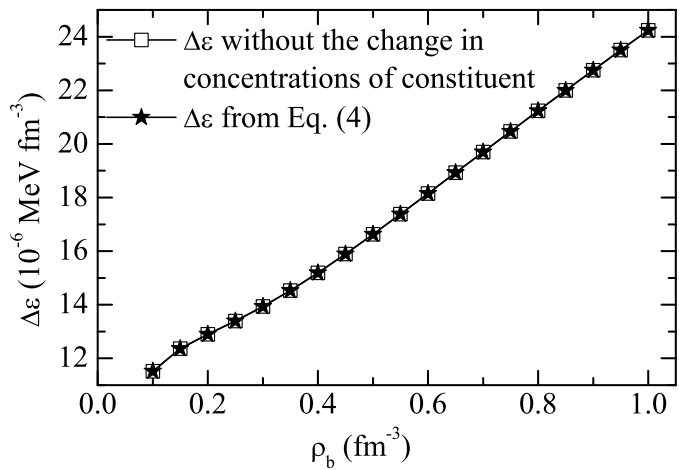}
\caption{Static thermal energy density $\Delta \varepsilon$ under
the temperature of $5\times 10^8$ K. The calculation without the
change in the concentration of each constitute is presented for
comparison.}\label{fig0}
\end{center}
\end{figure}

The stellar interior is assumed to be isothermal
described by a coordinate temperature (redshifted temperature),
since NSs are excellent conductors with quite high thermal
conductivities (Haensel et al. 2006). For the convenience of
calculation, we use the constant proper temperature (local
temperature) instead, which does not hinder us from discussing the
physics we are concerned with. Such finite temperature is not
expected to affect the stellar structure substantially. Accordingly,
the influence of temperature on the EOS acts as a small perturbation
in the present work. For the $\beta$-stable matter with a given
total baryon number density $\rho_b$, we derive the modifications of
the energy density and pressure due to the presence of temperature
$T$, which are given by
\begin{eqnarray}
\Delta \varepsilon  &=&m_{\sigma }^{2}\sigma \Delta \sigma
+g_{2}\sigma ^{2}\Delta \sigma +g_{3}\sigma ^{3}\Delta \sigma
+m_{\omega }^{2}\omega
_{0}\Delta \omega _{0}+m_{\rho }^{2}\rho _{30}\Delta \rho _{30} \notag \\
&&+6\Lambda _{v}g_{\rho }^{2}g_{\omega }^{2}\rho _{30}\omega
_{0}\left( \rho
_{30}\Delta \omega _{0}+\omega _{0}\Delta \rho _{30}\right) +\frac{\zeta }{2}%
g_{\omega }^{2}\omega _{0}^{3}\Delta \omega _{0}+ \notag \\
&&\underset{B=n,p}{\sum }\left[ \frac{1}{\pi ^{2}}\mu
_{B}^{2}\sqrt{\mu
_{B}^{2}-m^{\ast 2}}\Delta \mu _{B}+\frac{1}{6}\left( k_B T\right) ^{2}\frac{%
3\mu _{B}^{3}-2\mu _{B}m^{\ast 2}}{\sqrt{\mu _{B}^{2}-m^{\ast 2}}}\right] \notag \\
&&-\underset{B=n,p}{\sum }\frac{g_{\sigma }\Delta \sigma }{8\pi ^{2}}\bigg\{%
\frac{m^{\ast 2}-2\mu _{B}^{2}}{\sqrt{\mu _{n}^{2}-m^{\ast
2}}}m^{\ast }\mu
_{B}-2m^{\ast }\mu _{B}\sqrt{\mu _{B}^{2}-m^{\ast 2}}- \notag \\
&&4m^{\ast 3}\ln \frac{\mu _{B}+\sqrt{\mu _{B}^{2}-m^{\ast 2}}}{m^{\ast }}%
+m^{\ast 3}\frac{\mu _{B}^{2}+\mu _{B}\sqrt{\mu _{B}^{2}-m^{\ast
2}}}{\mu
_{B}^{2}-m^{\ast 2}+\mu _{B}\sqrt{\mu _{B}^{2}-m^{\ast 2}}}\bigg\} \notag \\
&&+\underset{l=e,\mu }{\sum }\left[ \frac{1}{\pi ^{2}}\mu
_{l}^{2}\sqrt{\mu
_{l}^{2}-m_{l}^{2}}\Delta \mu _{l}+\frac{1}{6}\left( k_B T\right) ^{2}\frac{%
3\mu _{l}^{3}-2\mu _{l}m_{l}^{2}}{\sqrt{\mu
_{l}^{2}-m_{l}^{2}}}\right], \label{BB1}
\end{eqnarray}
\begin{eqnarray}
\Delta P &=&-m_{\sigma }^{2}\sigma \Delta \sigma -g_{2}\sigma
^{2}\Delta \sigma -g_{3}\sigma ^{3}\Delta \sigma +m_{\omega
}^{2}\omega _{0}\Delta
\omega _{0}+m_{\rho }^{2}\rho _{30}\Delta \rho _{30}  \notag \\
&&+2\Lambda _{v}g_{\rho }^{2}g_{\omega }^{2}\rho _{30}\omega
_{0}\left(
\omega _{0}\Delta \rho _{30}+\rho _{30}\Delta \omega _{0}\right) +\frac{%
\zeta }{6}g_{\omega }^{2}\omega _{0}^{3}\Delta \omega _{0}+  \notag \\
&&\underset{B=n,p}{\sum }\left[ \frac{1}{3\pi ^{2}}\left( \mu
_{B}^{2}-m^{\ast 2}\right) ^{3/2}\Delta \mu _{B}+\frac{1}{6}\left(
k_B T\right)
^{2}\mu _{B}\sqrt{\mu _{B}^{2}-m^{\ast 2}}\right] -  \notag \\
&&\underset{B=n,p}{\sum }\frac{g_{\sigma }\Delta \sigma }{24\pi
^{2}} \bigg\{ \frac{5m^{\ast 2}-2\mu _{B}^{2}}{\sqrt{\mu
_{B}^{2}-m^{\ast 2}}}m^{\ast }\mu
_{B}+12m^{\ast 3}\ln \frac{\mu _{B}+\sqrt{\mu _{B}^{2}-m^{\ast 2}}}{m^{\ast }%
}  \notag \\
&&-10m^{\ast }\mu _{B}\sqrt{\mu _{B}^{2}-m^{\ast 2}}-3m^{\ast
3}\frac{\mu _{B}^{2}+\mu _{B}\sqrt{\mu _{B}^{2}-m^{\ast 2}}}{\mu
_{B}^{2}-m^{\ast 2}+\mu
_{B}\sqrt{\mu _{B}^{2}-m^{\ast 2}}}\bigg\}  \notag \\
&&+\underset{l=e,\mu }{\sum }\left[ \frac{1}{3\pi ^{2}}\left( \mu
_{l}^{2}-m_{l}^{2}\right) ^{3/2}\Delta \mu _{l}+\frac{1}{6}\left(
k_B T\right) ^{2}\mu _{l}\sqrt{\mu _{l}^{2}-m_{l}^{2}}\right] . \label{BB2}
\end{eqnarray}
Here $k_B$, $\mu$ and $m^*=m-g_{\sigma} \sigma$ are
the Boltzmann constant, zero-temperature chemical potential and
nucleonic Dirac effective mass, respectively. It should be stressed
that the $\mu_B$ for nucleons here is a translational chemical
potential defined as $\mu _{B}=\sqrt{k_{F,B}^{2}+m^{\ast 2}}$ for
the sake of derivation. $\Delta \mu _{B}$ and $\Delta \mu _{l}$ are
the changes in the chemical potentials of nucleons and leptons
respectively due to thermal effects. The changes in the meson fields
induced by thermal effects for a given baryon number density, namely
the $\Delta \sigma ,\Delta \omega _{0}$ and $\Delta \rho _{30}$, are
obtained with the conditions of $\beta$-stable and electric
neutrality of NS matter, which are respectively given as
\begin{eqnarray}
\Delta \sigma  &=&-g_{\sigma }m^{\ast }\left[ \underset{B=n,p}{\sum }\left(
\frac{1}{\pi ^{2}}\sqrt{\mu _{B}^{2}-m^{\ast 2}}\Delta \mu _{B}+\frac{\left(
kT\right) ^{2}}{6}\frac{\mu _{B}}{\sqrt{\mu _{B}^{2}-m^{\ast 2}}}\right) %
\right] \times \notag \\
&&\Bigg[(m_{\sigma }^{2}+2g_{2}\sigma +3g_{3}\sigma ^{2})+\frac{1}{2\pi ^{2}}%
g_{\sigma }^{2}m^{\ast 2}\Bigg(\frac{-\mu _{B}}{\sqrt{\mu _{B}^{2}-m^{\ast 2}}}+ \notag\\
&&2\ln \frac{m^{\ast }}{\mu _{B}+\sqrt{\mu _{B}^{2}-m^{\ast 2}}}+\frac{\mu
_{B}^{2}+\mu _{B}\sqrt{\mu _{B}^{2}-m^{\ast 2}}}{\mu _{B}^{2}-m^{\ast 2}+\mu
_{B}\sqrt{\mu _{B}^{2}-m^{\ast 2}}}\Bigg)\Bigg]^{-1}, \label{AA1}\\
\Delta \rho _{30} &=&\left[ m_{\rho }^{2}+2\Lambda _{v}g_{\rho
}^{2}g_{\omega }^{2}\omega _{0}^{2}-\frac{\left( 4\Lambda _{v}g_{\rho
}^{2}g_{\omega }^{2}\rho _{30}\omega _{0}\right) ^{2}}{m_{\omega
}^{2}+3c_{3}\omega _{0}^{2}+2\Lambda _{v}g_{\rho }^{2}g_{\omega }^{2}\rho
_{30}^{2}}+\frac{g_{\rho }^{2}x_{n}}{2}\right] ^{-1} \notag\\
&&\times \Bigg\{-g_{\rho }/2\left[ (\frac{x_{n}}{x_{p}}-1)\left( x_{e}+x_{\mu
}\right) +x_{n}\right] \Delta \mu _{e}- \notag\\
&&g_{\rho }/2\left[ (\frac{x_{n}}{x_{p}}-1)\left( y_{e}+y_{\mu
}-y_{p}\right) +y_{n}-y_{p}\right] \Bigg\},  \label{AA2}\\
\Delta \omega _{0}&=&\frac{-4\Lambda _{v}g_{\rho }^{2}g_{\omega }^{2}\rho
_{30}\omega _{0}}{m_{\omega }^{2}+3c_{3}\omega _{0}^{2}+2\Lambda _{v}g_{\rho
}^{2}g_{\omega }^{2}\rho _{30}^{2}}\Delta \rho _{30}, \label{AA3}\\
\Delta \mu _{e} &=&-\frac{\left( x_{p}+x_{n}\right) \left[
y_{e}+y_{\mu }-y_{p}\right] +x_{p}\left( y_{n}+y_{p}\right) }{\left(
x_{p}+x_{n}\right)
\left( x_{e}+x_{\mu }\right) +x_{p}x_{n}}\notag \\
&&-\frac{x_{p}x_{n}\left( g_{\rho }/2\right) 2}{\left(
x_{p}+x_{n}\right) \left( x_{e}+x_{\mu }\right) +x_{p}x_{n}}\Delta
\rho _{0},\nonumber\\ \nonumber
\end{eqnarray}
\begin{eqnarray}
\Delta \mu _{\mu} &=&\Delta \mu _{e },\notag \\
\Delta \mu _{p}&=&\frac{\left( x_{e}+x_{\mu }\right) \Delta \mu
_{e}+y_{e}+y_{\mu }-y_{p}}{x_{p}},\notag\\
\Delta \mu _{n} &=&\Delta \mu _{p}+\Delta \mu _{e}+\left( g_{\rho
}/2\right) 2\Delta \rho _{0},\notag \\
x_{B} &=&\frac{1}{\pi ^{2}}\sqrt{\mu _{B}^{2}-m^{\ast 2}}\mu
_{B},\quad B=n,p
\notag \\
y_{B} &=&\frac{g_{\sigma }\Delta \sigma }{\pi ^{2}}\sqrt{\mu
_{B}^{2}-m^{\ast 2}}m^{\ast }+\frac{\left( kT\right) ^{2}}{6}\frac{2\mu
_{B}^{2}-m^{\ast 2}}{\sqrt{\mu _{B}^{2}-m^{\ast 2}}},\notag \\
x_{l} &=&\frac{1}{\pi ^{2}}\sqrt{\mu _{l}^{2}-m_{l}^{2}}\mu _{l},\quad
l=e,\mu   \notag \\
y_{l} &=&\frac{\left( kT\right) ^{2}}{6}\frac{2\mu _{l}^{2}-m_{l}^{2}}{\sqrt{%
\mu _{l}^{2}-m_{l}^{2}}}.\notag
\end{eqnarray}
The above equations are solved in a self-consistent way. The static
thermal energy density under the temperature $T$ is $\Delta
\varepsilon$. The electrons in the neutron star crust may be
influenced considerably since the temperature in the crust is
sufficiently high for the motion of the electrons. Considering that
the pressure in the crust is primarily yielded by the electrons, the
effects of temperature on electrons are simply taken into account
via
\begin{equation}
\Delta \varepsilon =\frac{1}{6}\left( k_B T\right) ^{2}\mu
_{e}\sqrt{\mu _{e}^{2}-m_{e}^{2}},
\end{equation}%
\begin{equation}
\Delta P=\frac{\left( k_B T\right) ^{2}}{18}\sqrt{\mu
_{e}^{2}-m_{e}^{2}}\left( \frac{\mu _{e}^{2}+m_{e}^{2}}{\mu
_{e}}\right).
\end{equation}%
Due to the very small proportion of a star mass, the $\Delta
\varepsilon$ and $\Delta P$ from the crust contribute rather
insignificantly to the rearrangement of stellar structure.
Figure~\ref{fig0} displays the static thermal energy density $\Delta
\varepsilon$ as a function of density. Since the
chemical potential of each component changes slightly due to the
presence of temperature, the percentage of each component changes
accordingly as the result of chemical balance under a given baryon
number. The calculated $\Delta \varepsilon$ without the effect of
the change in the constitute concentrations, is also shown for
comparison, and we do not give the tedious formula here. It shows
that the two calculations are almost identical to each other,
indicating that the change in the concentrations due to the
temperature effect almost does not provide additional energy in the
present investigation. On the other hand, the
$\Delta \varepsilon$ can be calculated by $\Delta \varepsilon =\int \underset{i}{\sum }c_{i}dT$, where $c_{i}$ is the heat capacity per
unit volume for the species $i$. The $c_{i}$ is given as
$c_{i}=m_{i}^{\ast }k_{F,i}k_{B}^{2}T/(3\hbar ^{3})$ (Cumming et al.
2017), where $m_{i}^{\ast}$ is the Landau effective mass. The
nucleonic (non-relativistic) Landau mass is exactly equal to the Dirac mass
$m^*=m-g_{\sigma} \sigma$ here because the self-energy in the RMF
approach is momentum-independent. The obtained $\Delta \varepsilon$
with this method is also almost identical to the above results,
which indicates the correctness of our calculated thermal energy
density.

For the NS matter with temperature $T$ and baryon density $\rho_b$,
the energy density and pressure are written as $\varepsilon +\Delta
\varepsilon$ and $P+ \Delta P$, respectively. With these obtained
EOS for zero- or finite-temperature, the mass-versus-radius relation
and other relevant quantities of a spherically symmetric
non-rotating NS can be determined by solving the following TOV
equation (Oppenheimer \& Volkoff 1939) based on the general
relativity
\begin{eqnarray}
\frac{dP(r)}{dr} &=&-\frac{\varepsilon (r)m(r)}{r^{2}}\left( 1+\frac{P(r)}{%
\varepsilon (r)}\right) \left( 1+\frac{4\pi r^{3}P(r)}{m(r)}\right) \left( 1-%
\frac{2m(r)}{r}\right) ^{-1},  \notag \\
m(r) &=&\int_{0}^{r}4\pi r^{2}\varepsilon (r)dr,  \label{A0}
\end{eqnarray}
where $P(r)$ is the pressure of the star at distance $r$, $m(r)$ the
mass inside a sphere of radius $r$, and $c=G=1$. The radius $R$ and
mass $m(R)$ of a neutron star are obtained from the boundary
condition $P(R)=0$. Our EOS fulfills marginally the recent
observational maximum mass (Demorest et al. 2010; Antoniadis et al.
2013).

\begin{figure}[htbp]
\begin{center}
\includegraphics[width=0.6\textwidth]{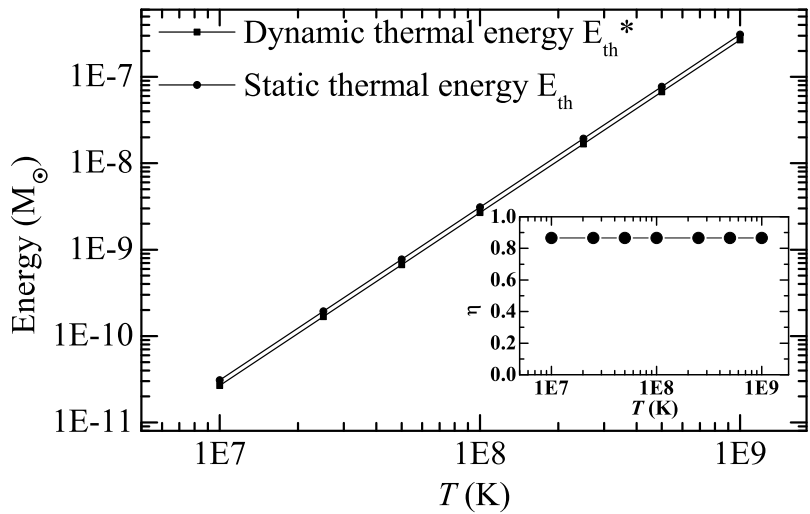}
\caption{The static thermal energy $E_{\text{th}}$ and the dynamic
thermal energy $E_{\text{th}}^*$ versus the core temperature $T$ for
the canonical NS. The inset displays the efficiency factor $\eta$ as
a function of $T$.}\label{fig1}
\end{center}
\end{figure}

\section{Dynamic treatment of NS thermal evolution}\label{intro}\noindent

NSs are born in supernova explosions, with internal temperature as
high as $\sim 10^{11}$ K, but cool down rapidly in about one minute
to become transparent to neutrinos (Haensel et al. 2006). Later, the
cooling is performed via two different channels--the neutrino
emission from the entire stellar body and the photon emission from
the stellar surface. The neutrino emission dominates the thermal
evolution for $t\lesssim 10^{5}$ years, while the photon emission
dominates later (Haensel et al. 2006; Yakovlev \& Pethick 2004). In
Newtonian framework, the energy balance equation for the NS cooling
is simply given as (Page et al. 2006)
\begin{equation}
\frac{dE_{\text{th}}}{dt}=C_{v}\frac{dT}{dt}=-L_{\nu }-L_{\gamma
}+H,\label{AA}
\end{equation}
where $T$ is the stellar internal temperature and $C_{v}$ is the
total heat capacity. The thermal energy $E_{\text{th}}$ is
dissipated by the neutrino emission (total luminosity $L_{\nu }$)
and photon emission (total luminosity $L_{\gamma }$). $H$ represents
all possible energy sources to heat the objects, such as the decay
of magnetic field energy stored in stars. The energy balance
equation offers an intuitive physical picture to describe the NS
cooling. Current simulations of thermal evolution including the
present work are usually on the basis of a general relativistic
formulation, and also carry out the heat transport inside the NSs,
where some robust program codes have already been established.

In fact, both the energy density and pressure of the dense matter
and hence the corresponding EOS, change as the temperature decreases
in stellar interior, and then the stellar structure reaches new
dynamic balances under different temperatures. In other words, the
structure of an isolated NS undergoes weak reorganizations
persistently during the cooling. In previous static descriptions,
the stellar structure is established before the cooling and not
modified thereafter. However, in our dynamic treatment, the NS
structure is modified continuously with temperature and hence the
time. Our starting point is to calculate the static
thermal energy $E_{\text{th}}=\int_{0}^{R }4\pi r^{2}\Delta
\varepsilon dr$ (Cumming et al. 2017) and the dynamic thermal energy
$E_{\text{th}}^*$. Note that the static thermal energy stored in a
NS is
$\int_{0}^{R }4\pi r^{2}\Delta \varepsilon /\sqrt{%
1-2m(r)/r}dr$, but that can be released actually is $\int_{0}^{R
}4\pi r^{2}\Delta \varepsilon dr$ in the static NS due to a strong
gravitational effect. The $E_{\text{th}}^*$ is exactly the
practical thermal energy that can be released during the cooling,
which is given by the total mass difference between finite
temperature $T$ and zero temperature cases, namely,
$E_{\text{th}}^*=m(T)-m(T=0)$. We must stress that,
in this process, the conserved quantity is total baryon number $n_b=\int_{0}^{\infty }4\pi r^{2}\rho_b /\sqrt{%
1-2m(r)/r}dr$, as a key constraint in the present calculations. As a
consequence, the effect induced by the continuous change of the NS
structure, is characterized by the difference between
$E_{\text{th}}$ and $E_{\text{th}}^*$. Owing to the rearrangement of
the stellar density distribution with the decrease of temperature,
the $E_{\text{th}}^*$ is not equal to the $E_{\text{th}}$ for a
given temperature.

\begin{figure}[htbp]
\begin{center}
\includegraphics[width=0.5\textwidth]{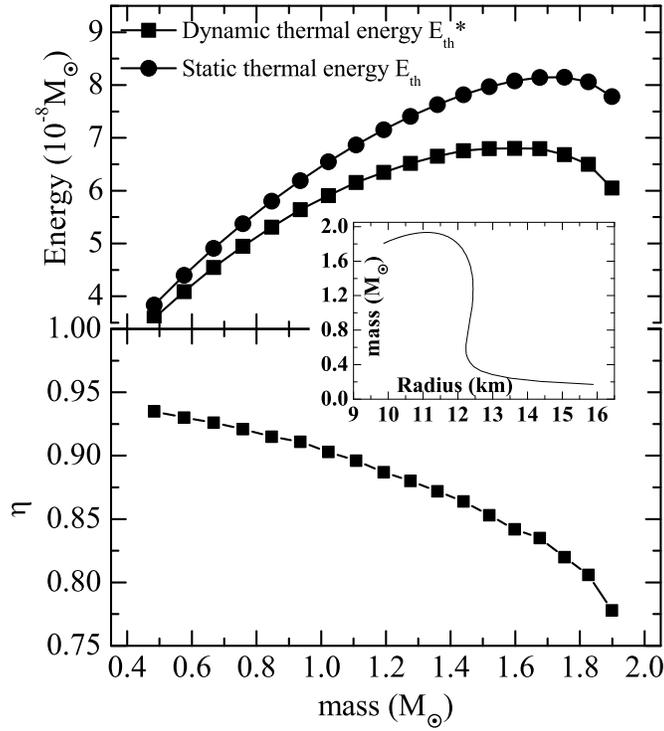}
\caption{(Upper panel) The static thermal energy $E_{\text{th}}$ and
the dynamic one $E_{\text{th}}^*$ for NSs with different masses. The
core temperature is chosen to be $5\times 10^8$ K. The lower panel
presents the corresponding efficiency factor $\eta$ as a function of
stellar mass. The inset displays the mass-radius relation of NSs
within the present EOS.}\label{fig2}
\end{center}
\end{figure}

The calculated $E_{\text{th}}$ and $E_{\text{th}}^*$ as a function
of interior temperature $T$ for a canonical NS with the mass of
$1.4M_{\odot}$, is displayed in Figure~\ref{fig1}. The stored
$E_{\text{th}}^*$ that will be lost during the cooling, is lower
than the static thermal energy \textcolor{blue}{$E_{\text{th}}=\int C_{v}dT$}. Importantly,
both of them are proportional to $T^2$, which allows us to introduce
an effective heat capacity $C_{v,\text{eff}}$ and hence the dynamic
thermal energy is written as \textcolor{blue}{$E_{\text{th}}^{\ast }=\frac{1}{2}C_{v,\text{eff}}T$}. It
is necessary to look for the relationship between $C_{v,\text{eff}}$
and $C_{v}$. Fortunately, the ratio of $\eta
=C_{v,\text{eff}}/C_{v}$, defined as the efficiency factor here, is
independent of temperature $T$, as shown in the inset of
Figure~\ref{fig1}. For example, $\eta =0.87$ for the canonical NS
with the IU-FSU interaction here, indicating that only $87\%$ of
static thermal energy is available for the NS cooling. As a result,
the traditional heat capacity $C_{v}$ should be replaced by the
effective one $C_{v,\text{eff}}$ or $\eta C_{v}$, to give a dynamic
description of NS cooling. Here we perform average with regard to
the whole star body for the efficiency factor $\eta$, i.e., $\eta$
is a constant everywhere.

For NSs with different masses, we calculate the $E_{\text{th}}$,
$E_{\text{th}}^*$ and the corresponding efficiency factor $\eta$,
and present the results in Figure~\ref{fig2}. For low mass NSs such
as $M=0.5M_{\odot}$, the $E_{\text{th}}$ is not much different from
the $E_{\text{th}}^*$, where the efficiency factor is as large as
$\eta =0.94$. However, with the increase of the star mass, the
difference between these two energies becomes larger and larger, and
hence the $\eta$ reduces gradually. For high mass NSs such as
$M=1.9M_{\odot}$, the difference is so substantial that $\eta$ is
merely 0.78; that is, $22\%$ of the static thermal energy is not
available. Figure~\ref{fig0} has already shown that the static
thermal energy density $\Delta \varepsilon$ (and hence the heat
capacity per volume) grows with density $\rho_b$, in consistent with
the calculations by Cumming {\it et al.} (2017). Yet, the NS radius
remains almost unchanged for the mass ranging from $0.4M_{\odot}$ to
$1.8M_{\odot}$, as shown in the inset of Figure~\ref{fig2}.
Therefore, the $E_{\text{th}}$ grows as the stellar mass increases
in this range. Beyond $1.8M_{\odot}$, NSs tend to be more compact
with increasing mass, leading to the $E_{\text{th}}$ decreases with
stellar mass as the result of these two competing effects. It is
well-known that the white dwarf is supported by the electron
degeneracy pressure, where gravity cannot compress it any more
because Pauli principle prevents it from complete collapse. However,
for NSs, it is the strong nuclear force among nucleons instead of
the nucleonic degeneracy pressure that contributes primarily to the
pressure. The alternation in the pressure $\Delta P$ induced by the
temperature stems primarily from the change of the $\sigma$ meson
field, where the $\sigma$ meson describes the intermediate range
attraction of nuclear force. Other meson fields change rather
insignificantly because the source terms in their respective motion
equations do not involve the temperature directly. The nonmonotonic
behavior of $E_{\text{th}}^*$ as a function of NS mass mainly
results from the temperature-induced nonmonotonic variation of the
$\sigma$ meson field versus density, yet, the $\eta$ decreases
monotonically with increasing mass.

The interior temperature of a NS is $10^7 \sim
10^9$ K usually, which is much lower than the corresponding Fermi
temperature of dense matter inside the star. Therefore, in order to
achieve a sufficient accuracy, we treat the effect of the
temperature on the EOS of dense matter as a tiny perturbation, and
then derive the analytical formulas (Eqs.~(\ref{BB1},\ref{BB2})) for
the changes in the total energy density and in pressure induced by
temperature, which is a pivotal step of the present study. By
setting up an initial central density, the TOV equation can be
integrated with the energy density $\varepsilon$ ($ \varepsilon
+\Delta \varepsilon$) and pressure $P$ ($P +\Delta P$) as inputs for
a zero- (finite-) temperature NS. The loop runs over the central
density of the finite-temperature NS until its total baryon mass
$m_b$ converges exactly to that of the zero-temperature one.
Actually, the interior structure of the zero-temperature NS, i.e.,
density and pressure distributions, is quite close to that of the
finite-temperature one because of $\Delta \varepsilon<< \varepsilon$
and $\Delta P << P$. Since the energy density $\varepsilon ^{\prime
}$ of the finite-temperature NS is given by the zero-temperature
energy density $\varepsilon ^{\prime }|_{T=0}$ plus the thermal
energy density $\Delta \varepsilon$, the dynamic thermal energy
$E_{\text{th}}^{\ast }=m(T)-m(T=0)$ is calculated by
\begin{eqnarray}
E_{\text{th}}^{\ast } &=&\int_{0}^{R}4\pi r^{2}(\varepsilon ^{\prime
}|_{T=0}+\Delta \varepsilon )dr-\int_{0}^{R}4\pi r^{2}\varepsilon
|_{T=0}dr \nonumber\\
&=&E_{\text{th}}+\int_{0}^{R}4\pi r^{2}(\varepsilon ^{\prime
}|_{T=0}-\varepsilon |_{T=0})dr,
\end{eqnarray}
in the present work based on $m_{b}(T)-m_{b}(T=0)=\int_{0}^{R}4\pi r^{2}dr(\rho _{b}^{\prime }/\sqrt{%
1-2m^{\prime }/r}-\rho _{b}/\sqrt{1-2m/r})=0$. Due to the
rearrangement of the stellar density distribution induced by
temperature, the zero-temperature energy density $\varepsilon
^{\prime }|_{T=0}$ of a finite-temperature NS is different from
$\varepsilon |_{T=0}$ of the corresponding zero-temperature one at
distance $r$, with $\varepsilon ^{\prime }|_{T=0}-\varepsilon
|_{T=0}<< \varepsilon$. The similar situation also applies to the
$m^{\prime }(r)$ and $m(r)$, $\rho _{b}^{\prime }$ and $\rho _{b}$.
The difference between the $E_{\text{th}}^{\ast }$ and
$E_{\text{th}}$ is $\Delta E_{\text{th}}=\int_{0}^{R}4\pi
r^{2}(\varepsilon ^{\prime }|_{T=0}-\varepsilon |_{T=0})dr$, which
is exactly the change of the zero-temperature total energy (internal
energy plus gravitational potential energy) due to the rearrangement
of stellar structure. We tested the calculation and find that the
results almost do not rely on a step size (if it is reasonably
small) of solving TOV equation, which supports the validity of the
present calculation.

\begin{figure}[htbp]
\begin{center}
\includegraphics[width=0.5\textwidth]{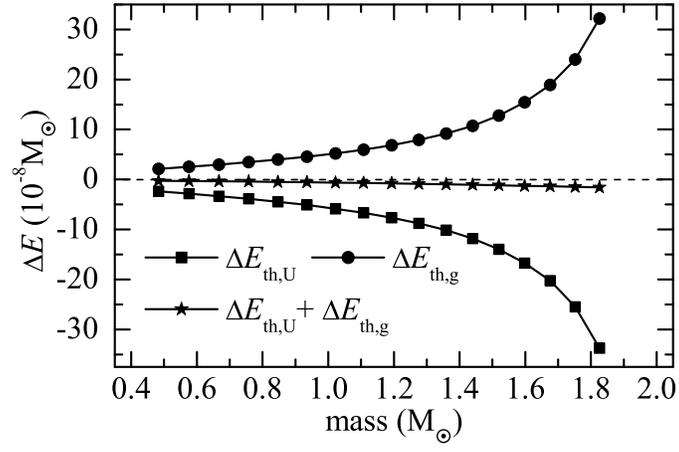}
\caption{Calculated $\Delta E_{\text{th, U}}$ and $\Delta
E_{\text{th, g}}$ as a function of the stellar mass. The core
temperature is chosen to be $5\times 10^8$ K.} \label{fig25}
\end{center}
\end{figure}

Furthermore, the $\Delta E_{\text{th}}$ can be
decomposed into two parts:
\begin{eqnarray}
\Delta E_{\text{th}} &=&\int_{0}^{R}4\pi r^{2}\left(
\frac{\varepsilon
^{\prime }|_{T=0}}{\sqrt{1-2m^{\prime }/r}}-\frac{\varepsilon |_{T=0}}{\sqrt{%
1-2m/r}}\right) dr \nonumber \\
&&+\int_{0}^{R}4\pi r^{2}\left( \varepsilon ^{\prime }|_{T=0}\frac{\sqrt{%
1-2m^{\prime }/r}-1}{\sqrt{1-2m^{\prime }/r}}-\varepsilon |_{T=0}\frac{\sqrt{%
1-2m/r}-1}{\sqrt{1-2m/r}}\right) dr,
\end{eqnarray}
where the first and second integrals denote the change of the
zero-temperature internal energy and of the gravitational potential
energy induced by temperature, marked as $\Delta E_{\text{th, U}}$
and $\Delta E_{\text{th, g}}$ respectively. Figure~\ref{fig25}
displays the calculated $\Delta E_{\text{th, U}}$ and $\Delta
E_{\text{th, g}}$ versus stellar mass. The central density of the
finite-temperature NS is slightly lower than that of
zero-temperature one. That is, due to the presence of temperature,
the dense matter deep inside the star moves outward. This leads to
the decrease of zero-temperature internal energy but the increase of
the gravitational potential energy, and such changes are mainly
subject to the change of the compactness parameter $2m/r$. Because
the thermal energy density $\Delta \varepsilon$ becomes larger and
larger with increasing density as shown in Figure~\ref{fig0}, the
difference between the compactness parameter $2m^{\prime }/r$ of
finite-temperature NS and $2m/r$ of the corresponding
zero-temperature one, i.e., $\sim 2E_{\text{th}}(r)/r$ with
$E_{\text{th}}(r)=\int_{0}^{r }4\pi r^{2}\Delta \varepsilon dr$,
grows with NS mass. Therefore, the absolute values of both $\Delta
E_{\text{th, U}}$ and $\Delta E_{\text{th, g}}$ grow with stellar
mass. The net temperature effect is a result of these two competing
trends.

\begin{table*}[h]
\label{table1} \caption{A ¡°family¡± of IU-FSU interaction. $L=3\rho
\partial E_{\text{sym}}(\rho )/\partial \rho |_{\rho _{0}}$ is the slope parameter of the symmetry energy $E_{\text{sym}}$ at the
saturation density $\rho_{0}$.}
\begin{tabular}{cccccccc}
\hline
$\Lambda_{\text{V}}$  & $g_{\rho}^{2}$ & $L$(MeV) &\quad \quad &\quad \quad& $\Lambda_{\text{V}}$  & $g_{\rho}^{2}$ & $L$(MeV)  \\
\hline
$0.00$ & 84.4175 & 119   &\quad \quad& \quad \quad& $0.01$ & 95.7143 & 91   \\

$0.02$ & 110.5015  & 72  &\quad \quad& \quad \quad&  $0.03$ & 130.6928 & 60  \\

$0.04$ & 159.9126  & 51  &\quad \quad& \quad \quad& $0.05$  & 205.9605  & 45  \\
\hline
\end{tabular}
\end{table*}

\begin{figure}[htbp]
\begin{center}
\includegraphics[width=0.5\textwidth]{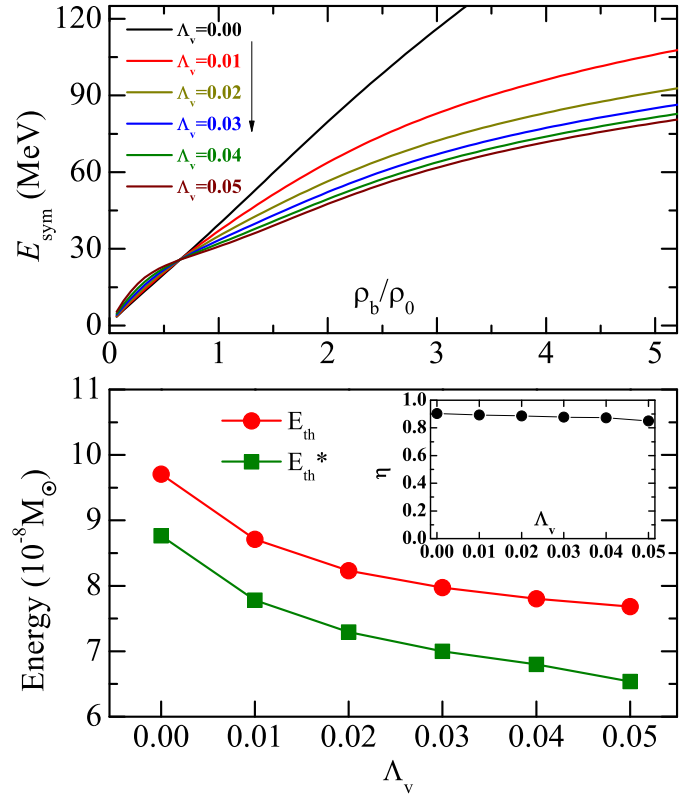}
\caption{(Upper panel) The symmetry energy as a function of density
(in units of the saturation density $\rho_0$) for the family of
IU-FSU interactions. (Lower panel) The calculated static thermal
energy $E_{\text{th}}$ and the dynamic one $E_{\text{th}}^*$ for
canonical NSs under different behaviors of symmetry energy. The core
temperature is $5\times 10^8$ K. The inset presents the
corresponding efficiency factor $\eta$. (A color version of this figure is available in
the online journal.)}\label{fig3}
\end{center}
\end{figure}

Nowadays the properties of dense matter at supersaturation densities
remain uncertain due to poor knowledge of the nuclear forces and the
difficulty of solving the many-body problem. Therefore, it is
necessary to test the uncertainty of the $E_{\text{th}}^*$ and
$\eta$ caused by different interactions. The IU-FSU interaction used
here describes well the EOS of symmetric matter (Fattoyev et al.
2010), but the symmetry energy which characterizes the
isospin-dependent part of the EOS of asymmetric nuclear matter
(Baran et al. 2005; Steiner et al. 2005; Lattimer \& Prakash 2007;
Li et al. 2008) is not yet well determined at high densities. To
explore the sensitivity of $E_{\text{th}}^*$ and $\eta$ to the
symmetry energy, we build a ¡°family¡± of IU-FSU interactions by
adjusting the isovector parameters $\Lambda_{\text{V}}$ and
$g_{\rho}$ in such a way that the value of the symmetry energy
remains fixed at 25.70 MeV at a baryon density of $\rho_b =0.1$
fm$^{-3}$ within a procedure as in the work of Piekarewicz (2011).
The parameters $\Lambda_{\text{V}}$ and $g_{\rho}^{2}$ are listed in
Table I with the slope $L=3\rho
\partial E_{\text{sym}}(\rho )/\partial \rho |_{\rho _{0}}$ at the
saturation density $\rho_{0}$, which could also be employed to
investigate the effects of the density-dependent symmetry energy in
some subjects in the future. To illustrate the behavior of the
mean-field interactions, we display in the upper panel of
Figure~\ref{fig3} the symmetry energy predicted by all of these
models, providing a stiff symmetry energy
($\Lambda_{\text{V}}=0.00$) to a soft one
($\Lambda_{\text{V}}=0.05$). The convergence of all interactions at
a density of $\rho_b =0.1$ fm$^{-3}$ is clearly discernible, and are
divergent visibly at high densities. For instance, the stiff
symmetry energy with $\Lambda_{\text{V}}=0.00$ gives
$E_{\text{sym}}(3\rho_0)=116$ MeV while the relatively soft one with
$\Lambda_{\text{V}}=0.05$ gives $E_{\text{sym}}(3\rho_0)=62$ MeV,
where $3\rho_0$ is close to the central density of a canonical NS.
The lower panel of Figure~\ref{fig3} illustrates the $E_{\text{th}}$
and $E_{\text{th}}^*$ under the different behavior of the symmetry
energy. A stiffer symmetry energy gives a larger $E_{\text{th}}$ and
$E_{\text{th}}^*$. From $\Lambda_{\text{V}}=0.00$ to
$\Lambda_{\text{V}}=0.05$, the $E_{\text{th}}$ and $E_{\text{th}}^*$
reduce by $21\%$ and $25\%$, respectively. Yet, on the whole, the
effect of the symmetry energy on the efficiency factor $\eta$ is not
so intense because the effect of symmetry energy plays a similar
role for the $E_{\text{th}}$ and $E_{\text{th}}^*$, as exhibited in
the inset. The depressed model dependence enhances the reliability
of the presently obtained $\eta$.

\begin{figure}[htbp]
\begin{center}
\includegraphics[width=0.6\textwidth]{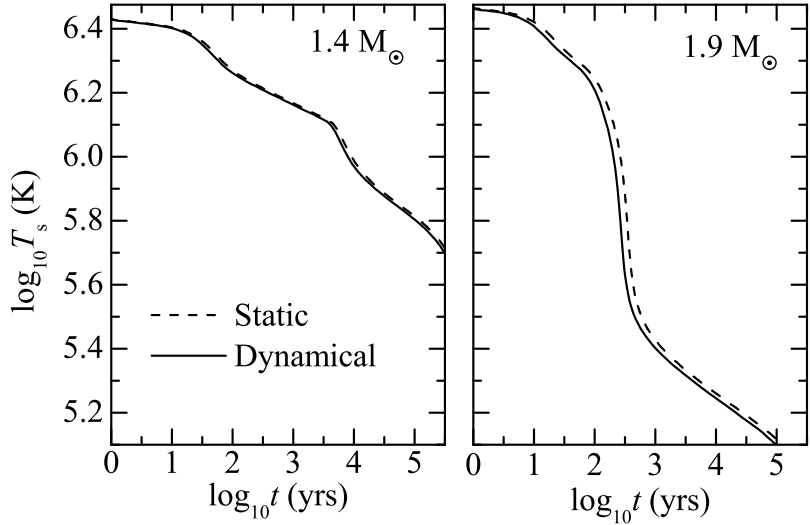}
\caption{Cooling curves of a $1.4M_{\odot}$ and a $1.9M_{\odot}$
NSs. The calculations are carried out with the static (traditional)
and dynamic (present) treatments. The stellar structure is also
built by employing the RMF approach with the IU-FSU interaction, and
the effective mass is taken from the APR EOS (Akmal et al.
1998).}\label{fig4}
\end{center}
\end{figure}

The central task of NS cooling theory is to calculate the cooling
curve, i.e., the surface temperature as a function of age. To show
the difference between the dynamic and static treatments of the NS
cooling, we compute the cooling curves for a canonical
($1.4M_{\odot}$) and a large mass ($1.9M_{\odot}$) NSs based on the
publicly available code NSCool written by D. Page $^{[1]}$
\footnotetext[1]{http://www.astroscu.unam.mx/neutrones/home.html}.
The minimal cooling paradigm is used, i.e., without charge-meson
condensate and exotic degrees of freedom. Yet, for large mass NSs,
the direct Urca process is open within the IU-FSU interaction since
the matter density in the core could exceed the threshold. Although
such a process is believed to enhance the NS cooling most
efficiently, it is not included in standard cooling scenario for
canonical NSs. The results are shown in Figure~\ref{fig4}. Because
the dynamic thermal energy is less than the static one, the NS
cooling is found to be faster in dynamic treatment than that in
static treatment. For the canonical NS and the large mass NS, the
two approaches do not exhibit considerable difference, indicating
that the static description is a good approximation. A reliable NS
cooling theory is indispensable to help one to explore the knowledge
of stellar interior. However, owing to the complexity of neutron
star physics (such as anisotropic magnetic field, and composition of
stellar envelope), it still has a long way to go to establish an
ultimate cooling theory. On the other hand, much more observational
data are required to constrain in turn the cooling theory.

\section{Summary}\label{intro}\noindent
Although the NS cooling theory was established more than half a
century ago, it is based on the assumption that the stellar
structure is static. For a given NS with fixed baryon number, its
structure undergoes weak reconfiguration continuously as the
temperature decreases gradually, which is attributed to the
alteration in EOS of the $\beta$-stable NS matter. As a crucial step
in the present study, we have derived the EOS of dense matter at
finite temperature $T$ with a perturbation method, where $T$ is much
lower than the Fermi temperature. Correspondingly, the the concept
of dynamic thermal energy is introduced to distinguish it from the
previous static thermal energy. The dynamic thermal energy is found
to be less than the static one. Therefore, a part of static thermal
energy with a given core temperature cannot be released during the
cooling. Thus, we defined a temperature-independent efficiency
factor $\eta$ to characterize it, which is subject to the NS mass
but is not sensitive to the different behavior of symmetry energy.
In the dynamic description of NS thermal evolution that the change
of stellar structure is included in the process of cooling, one just
needs to replace the heat capacity $C_{v}$ in previously static
approach by the effective one $C_{v,\text{eff}}=\eta C_{v}$.
Finally, based on the above discussions, we computed the cooling
curves with both the static and dynamic treatments. The cooling
turns out to be slightly faster in our dynamic treatment than that
in the static one. The static description of cooling can be regarded
as a reasonable approximation for NSs with different masses.
Therefore, our work not only examines whether the traditional static
description of cooling is valid or not, but also deepens our
understanding of the NS cooling.

The weak rearrangement of stellar structure discussed above stems
from the decrease of temperature where the total baryon number of a
star is conserved. Actually, the rearrangement can be achieved
through many other avenues, such as the spin-down, magnetic-field
decay, and possible phase transitions, which should improve our
understanding of reheating mechanism in NS cooling and is perhaps
responsible for some intriguing features of NSs.

\section*{Acknowledgement}
\label{intro}\noindent J. M. Dong gratefully acknowledges the
support of K. C. Wong Education Foundation. This work was supported
by the National Natural Science Foundation of China under Grants
Nos. 11435014, 11775276, 11405223, 11675265, and 11575112, by the
973 Program of China under Grant No. 2013CB834401 and No.
2013CB834405, by the National Key Program for S\&T Research and
Development (No. 2016YFA0400501 and No. 2016YFA0400502), by the
Knowledge Innovation Project (KJCX2-EW-N01) of Chinese Academy of
Sciences, by the Funds for Creative Research Groups of China under
Grant No. 11321064, and by the Youth Innovation Promotion
Association of Chinese Academy of Sciences.


\end{document}